\documentclass[11pt]{amsart}
\usepackage[english]{babel}

\usepackage{enumerate}
\usepackage{a4wide,amssymb,amsmath}
\newcommand{\be}{\begin{equation}} \newcommand{\ee}{\end{equation}}
\newcommand{\bea}{\begin{eqnarray*}}
\newcommand{\eea}{\end{eqnarray*}} \newcommand{\beq}{\begin{eqnarray}}
\newcommand{\eeq}{\end{eqnarray}}
\newcommand{\RR}{\mathbb{R}}
\newcommand{\NN}{\mathbb{N}}

\newcommand{\ZZ}{\mathbb{Z}}
\newcommand{\EE}{\mathbb{E}}
\newcommand{\PP}{\mathbb{P}}

\newcommand{\cE}{\mathcal{E}}
\newcommand{\cF}{\mathcal{F}}

\newcommand{\cH}{\mathcal{H}}

\newcommand{\LL}{{\Lambda_L}}
\renewcommand{\r}{\right}
\renewcommand{\l}{\left}
\newcommand{\la}{\langle}
\newcommand{\ra}{\rangle}

\newcommand{\supp}{\mathop{\mathrm{supp}}}

\newcommand{\Tr}{\mathop{\mathrm{Tr}}}
\newcommand{\per}{\mathrm{per}}

\newcommand{\tl}{\tilde \lambda}

\newcommand{\tV}{\tilde V}
\newcommand{\tH}{\tilde H}

\newtheorem{thm}{Theorem}
\newtheorem{lem}[thm]{Lemma}

\newtheorem{cor}[thm]{Corollary}
\theoremstyle{definition}

\newtheorem{rem}{\textsl{Step}}
\newtheorem{ex}[thm]{Example}
\newenvironment{exm}{\begin{ex}}{\hfill\qed\end{ex}}
\newcommand{\Hm}[1]{\leavevmode{\marginpar{\tiny%
$\hbox to 0mm{\hspace*{-0.5mm}$\leftarrow$\hss}%
\vcenter{\vrule depth 0.1mm height 0.1mm width \the\marginparwidth}%
\hbox to 0mm{\hss$\rightarrow$\hspace*{-0.5mm}}$\\\relax\raggedright #1}}}

\begin{document}
\title[Lifshitz tails for the breather model]{Lifshitz tails
for a class of Schr\"odinger operators  with random breather-type potential}
\author[W.~Kirsch]{Werner Kirsch}
\address[W.~K.]{Fakult\"at f\"ur
Mathematik und Informatik \\FernUniversit\"at Hagen, Germany}
\urladdr{www.fernuni-hagen.de/WTHEORIE/}

\author[I.~Veseli\'c]{Ivan Veseli\'c}
\address[I.~V.]{Fakult\"at f\"ur Mathematik, 09107  TU-Chemnitz, Germany   }
\urladdr{http://www.tu-chemnitz.de/mathematik/stochastik}
\thanks{\copyright 2008 by the author. Faithful reproduction of this article, is permitted for non-commercial purposes. 
This work has been partially supported by the DFG within the Emmy-Noether-Project 
``Spectral properties of random Schr\"odinger operators and random operators on manifolds and graphs''. }

\date{ \jobname.tex}

\keywords{random Schr\"odinger operators, integrated density of states, Lifshitz tails, breather model, non-linear randomness}
\subjclass[2000]{35J10; 82B44}

\begin{abstract}
We derive bounds on the integrated density of states
for a class of Schr\"odinger operators with a random potential.
The potential depends on a sequence of random variables, not necessarily in a linear way.
An example of such a random Schr\"odinger operator is the breather model,
as introduced by Combes, Hislop and Mourre.
For these models we show that the integrated density of states near the bottom
of the spectrum behaves according to the so called Lifshitz asymptotics.
This result can be used to prove Anderson localization in certain energy/disorder regimes.
\end{abstract}
\maketitle

\section{Introduction, model and result}
\label{s-intro}
In this paper we study spectral properties of certain
Schr\"odinger operators with random potential.
The spectral theory of such operators has been studied since
the eighties in the mathematical literature and
there are several monographs devoted to this topic,
see e.g.~\cite{CyconFKS-87,CarmonaL-90,PasturF-92,Stollmann-01}.
Certain spectral features,
like the non-randomness of the spectral components and the integrated density of states, are
shared by a wide variety of models under mild ergodicity and regularity assumptions.
However specific characteristics ---  like the existence of a certain spectral type ---
depend on the concrete model at hand.

Our aim is to establish for a class of random Schr\"odinger operators
the Lifshitz asymptotics of the integrated density of states (in the sequel abbreviated by IDS).
Spectral edges at which the IDS exhibits Lifshitz tails are called fluctuation boundaries (of the spectrum). Based on physical intuition one expects that ergodic random Hamilton operators exhibit
spectral localization in a neighbourhood of a fluctuation boundary. Here spectral localization 
means that in the relevant energy interval the spectrum is pure point, and the continuous spectral component is absent, almost surely. 
Rigorous proofs of localization oftentimes rely on the estimates on the finite volume approximation of the IDS implied by the Lifshitz asymptotics. 

We define now the class of operators considered in this paper and thereafter present our results.  
They concern random Schr\"odinger operators $H_\omega$ of the following type 
\be\label{def:Operator}
H_\omega~=~H_0\,+\,V_\per\,+\,V_\omega
\ee
where $H_0=-\Delta$ denotes the Laplacian on $L^2(\mathbb{R}^d)$, $V_\per$ is a ($\ZZ^d$-)periodic potential
and $V_\omega$ is a random potential of the form

\be
\label{e-W}
V_\omega(x) := \sum_{k\in\ZZ^d} u(\lambda_k (\omega) , x-k) .
\ee
Here $ \lambda_k \colon \Omega \to [\lambda_-,\lambda_+], k\in\ZZ^d$
is a  collection of non-trivial, independent, identically distributed random variables
on the probability space $(\Omega,\cF, \PP)$. The distribution of $\lambda_0$ is denoted
by $\mu$ and we assume $\inf \supp \mu = \lambda_-$. 
The function $u$ is called \emph{single site potential}.
\smallskip

Throughout this paper we will make the following 
\smallskip

\noindent\textbf{Assumptions:}\\
\textsl{Periodic Potential:}
The potential $V_\per$ is $\ZZ^d$-periodic and locally in $L^p(\RR^d)$ for some $p>d$.
\medskip

\noindent\textsl{Random Potential:}
The single site potential $u \colon \RR \times \RR^d \to \RR$  is jointly measurable
and satisfies the following:
{\renewcommand{\labelenumi}{(\roman{enumi})}
\begin{enumerate}
\item For all $\lambda \in [\lambda_-,\lambda_+]$ we have
\begin{equation*}
\supp u(\lambda , \cdot) \subset \Lambda_1 := \left [-\frac{1}{2},\frac{1}{2}\right]^d
\end{equation*}
\item \label{a-kappa1} We have for all $\lambda \in [\lambda_-,\lambda_+]$ 
\begin{equation*}
\frac{\partial}{\partial\lambda}u(\lambda,\cdot) \in L^\infty(\Lambda_1)
\quad \text{ with } \quad
\kappa_1 := \sup_{x \in \RR^d} \sup_{\lambda \in [\lambda_-,\lambda_+]}
\left|\frac{\partial}{\partial\lambda}u(\lambda,x) \right| < \infty
\end{equation*}

\item\label{a-pos1} For all $ x \in \RR^d$ and $ \lambda \in [\lambda_-,\lambda_+]$ we have
\begin{equation*}
\frac{\partial u}{\partial \lambda} (\lambda, x) \ge 0
\end{equation*}

\item \label{a-pos2} There exist $\epsilon_1, \epsilon_2 > 0$ such that for all
$\lambda \in [\lambda_-,\lambda_-+\epsilon_2]$ we have
\begin{equation*}
\frac{d}{d\lambda} \int_{\RR^d} dx \, u(\lambda, x) \in [\epsilon_1,1/\epsilon_1]
\end{equation*}
\item There exist $\alpha, \kappa >0$ such that
 for all $\epsilon  \le \epsilon_2$
\bea
\mu\l([\lambda_-,\lambda_-+\epsilon)\r)~\geq \alpha \, \epsilon^\kappa
\eea
\end{enumerate}}
Note that due to the positivity assumption (ii) we have actually 
$$\kappa_1 = \sup_{x \in \RR^d} \sup_{\lambda \in [\lambda_-,\lambda_+]}
\frac{\partial}{\partial\lambda}u(\lambda,x)
$$

A single site potential which stisfies conditions (i) - (iv) will be called \emph{monotone in the randomness}.
Note that the randomness enters the potential \eqref{e-W} via a field of random
variables $\lambda_k, k\in\ZZ^d$, not necessarily in a linear way.
A random potential of the form \eqref{e-W} with a single site potential satisfying the above
Assumptions gives rise to a \emph{metrically transitive} or \emph{ergodic operator},
see e.g.~\cite{Kirsch-89a} or \cite{PasturF-92} for the definition.  This implies that
there is a subset $\Sigma$ of the real line  such that the spectrum of $H_\omega$
coincides with $\Sigma$ almost surely, and that there is a well defined IDS for the family 
$H_\omega, \omega\in \Omega$, see below for details.
\smallskip

Condition (iii) ensures that for all $\omega \in \Omega, x\in \RR^d, k\in \ZZ^d$ we have
\be
\label{e-monotone}
u(\lambda_k(\omega), x) \ge u (\lambda_-,x)
\ee
This enables us to write the potential part of the operator in a standardized way.
If we set
\bea
V_0(x) := \sum_{k\in\ZZ^d} u(\lambda_-, x-k)\quad\text{and}\quad \tilde u ( \lambda, x ) :=u ( \lambda, x ) -u ( \lambda_-, x )
\eea
then $V_0$ is $\ZZ^d$-periodic and $\tilde u$ satisfies the same assumptions as the original single site potential $u$. Moreover $\tilde{u}(\lambda,x)\geq 0$ and $\tilde{u}(\lambda_-,x)= 0$. 
Thus we may subsume $V_0$ into the potential $V_\per$ using the relation $V_\per\,+\,V_\omega = (V_\per+ V_0) \,+\,(V_\omega- V_0)  $. Consequently, we may and will assume from now on without loss of generality that
\be\label{a-utilde}
u(\lambda,x)\geq 0\quad\text{and}\quad u(\lambda_-,x)= 0\,.
\ee
It is easy to see that in this case
$E_0:=\inf\sigma(H_\per)$ equals $\inf\sigma(H_\omega)$ almost surely.

\bigskip

\begin{exm}
\label{x-alloy}
If we set in \eqref{e-W}
$u(\lambda , x-k)= 
\lambda \, f( x-k)$
we obtain an \emph{alloy type potential}
\be\label{e-alloy}
V_\omega(x) := \sum_{k \in \ZZ^d} \lambda_k(\omega) \ f(x-k)
\ee
Such random potentials have been thoroughly studied before
in the context of the Lifshitz asymptotics of the integrated density of states and localization,
see e.g.~\cite{KirschM-83a,MartinelliH-84,KirschS-86,Mezincescu-87,CombesH-94b,
Klopp-95a,Kirsch-96,KirschSS-98a,Stollmann-99a,DamanikS-01,Stollmann-01,GerminetK-01a,GerminetK-04,AizenmanENSS-06}.
If $f$ is non-negative and sufficiently regular
the resulting single site potential $u$ is monotone in the randomness.
For such alloy type models the results we are aiming at
are by now well understood, therefore we will not elaborate on them further.
\end{exm}
\begin{exm}
The main example which motivated this paper was introduced in \cite{CombesHM-96}. For this model we set
\bea
u(\lambda, x) = -f(\lambda \, x)
\eea
The resulting stochastic field
\be\label{e-breatherpotential}
V_\omega(x) := \sum_{k \in \ZZ^d}  \ -\,f(\lambda_k(\omega)(x-k))
\ee
is called random \emph{breather-type potential}, cf.~\cite{CombesHM-96,CombesHN-01}.
If we assume for the function $f$ that
\begin{align}
\label{e-fC2}
\supp f \subset \Lambda_{\lambda_-}, \quad f \in  C_0^1(\RR^d\setminus \{0\})
\\
\label{e-frepulsive}
L^\infty(\RR^d) \ni g(x):=  -x \cdot (\nabla f)(x)\ge 0 \text{ and not identically vanishing}
\end{align}
then the potential $u \colon \RR \times \RR^d \to \RR$  is monotone in the randomness.
Inequality \eqref{e-frepulsive} is called the \emph{repulsivity property} of $f$.
\end{exm}
In order to formulate our main result we introduce some more notation.
For a selfadjoint operator $A$ and a Borel set $J\subset\RR$ we denote the associated
spectral projection by $\chi_J(A)$.
Let $\Lambda_L(j):= [-L/2, L/2]^d+j\subset\RR^d$ be a cube of side length $L$ 
centered at $j\in\ZZ^d$. We write $\Lambda_L$ for $\Lambda_L(0)$ and denote by $ \chi_{\Lambda_L}$ the characteristic function of this set.
One possible way to define the IDS $N\colon \RR \to \RR$ is the following 
trace per unit volume formula
\[
 N(E) := \EE \left \{\Tr [\chi_{\Lambda_1}\,  \chi_{]-\infty,E]}(H_\omega)  ]  \right	\}
\]
The almost sure infimum $E_0$ of the spectrum of the operator $H_\omega$ coincides with the 
energy $\inf \{E\in \RR \mid N(E) >0\} $. For more details about the IDS we refer to the
surveys \cite{KirschM-07} or \cite{Veselic-07b} and references given there.


The follwing result on the asymptotics of the IDS at the bottom of the spectrum tells us that 
it behaves roughly like $N(E) \sim \exp \big(-const.\, (E-E_0)^{-d/2}\big)$ for $E-E_0$ positive and small.

\begin{thm}[Lifshitz Tails]
\label{t-Lifshitz}
Let $H_\omega, \omega  \in \Omega$ be a random operator with potential \eqref{e-W}
satisfying the above assumptions.  Then
\begin{equation}
    \label{e-LifTaildef}
    \lim_{E \searrow E_0} \frac{\log |\log N(E)|}{\log (E-E_0)} = -\frac{d}{2}
\end{equation}
\end{thm}
\bigskip
This result can be used as a tool in a proof of spectral localization for breather-type models. In fact, the localization proof based on 
multiscale analysis usually requires two main ingredients, the Wegner estimate and an initial scale estimate.
A Wegner estimate for breather-type models was given in \cite{CombesHM-96} (see that paper for precise assumptions
and \cite{CombesHN-01} for related results).
Those authors also prove localization under certain assumptions on the disorder and the energy. One can replace their initial scale
estimate using Theorem \ref{t-Lifshitz} to obtain localization for small energies for breather-type models. We will not give details
here.

Let us note that recently there has been a number of papers devoted to Lifshitz tails  for models which depend
non-monotonously on the randomness, cf.~\cite{BakerLM-08,Fukushima-09a,KloppN-09a}.
 To treat such models one needs to use different methods than ours.

Let us make a comment on the difference between our proof of Lifshitz tails and the one for the standard alloy type model.
The basic strategy of proof is the same, but since we are dealing here with a non-linear dependece on the randomness, we 
introduce a new family of (non-linearly) mapped random variables, which correspond to local energy contributions. 
Then we are in the position to make use of Temple's inequality similarly as in the case of alloy type potentials. 
To be able to control the  relation of the second moment (of the energy) to the first moment we need to linearize it. 
This linearisation is one of the instances where we need the differentiability of the single site potentials with respect to the parameter $\lambda$
and sufficient control on the derivative.

\bigskip
As usual we prove Theorem \ref{t-Lifshitz} by giving an upper and a lower bound on $N$, respectively on the limit in Eq.~\eqref{e-LifTaildef}.
The next Section \ref{s-bc} is of preparatory nature where we discuss boundary conditions and corresponding bounds on the integrated density of states.
The subsequent Section \ref{s-proof} contains the proof of the upper bound on the IDS. The lower bound is given in the final Section \ref{s-lower}.

\section{Boundary conditions}\label{s-bc}

In this section we discuss boundary conditions for operators on cubes $\LL$. 
For the upper bound on the IDS we will use operators $H_0^{L,\rho}$ on $L^2(\LL)$ with appropriate `mixed' boundary conditions,
formulated in terms of a bounded function $\rho$  on the boundary $\partial\LL$ of $\LL$.
We say that a smooth function $\phi$ 
obeys the $\rho-$boundary conditions if $\rho\, \phi = - n\cdot \nabla\phi$ on $\partial\LL$.
Here $n\cdot\nabla$ denotes the outer normal derivative at $\partial\LL$.

In a rigorous way we
define $H_0^{L,\rho}$ as the operator associated to the sesquilinear form
\bea
(\phi_1,\phi_2)\mapsto
\int_{\Lambda_L}\,\overline{\nabla \phi_1(x)}\,\nabla \phi_2(x)\,dx\;+\;\int_{\partial\LL}\,\rho(x)\, \overline{\phi_1(x)}\phi_2(x)\,dx
\eea
with the Sobolev space $\cH^1(\LL)$ as its form domain. Here we use the same notation for a function $\phi$ on the cube $\Lambda_L$
and its trace on $\partial \Lambda_L$. Note that the second term of the sesquilinear form is well defined since the trace of $\phi$ is in $L^2(\partial \Lambda_L)$.
The Neumann operator $H_0^{L,N}$ is given by the choice $\rho\equiv0$, Dirichlet boundary conditions are
formally given by $\rho\equiv\infty$. The Laplacian with Dirichlet boundary conditions  $H_0^{L,D}$ is rigorously defined through the form
\bea
(\phi_1,\phi_2)\mapsto\int_{\Lambda_L}\,\overline{\nabla \phi_1(x)}\,\nabla \phi_2(x)\,dx
\eea
on $\cH_0^1(\LL)$.

We will also need the operator $H_0^{L,P}$, the Hamiltonian with periodic boundary conditions at $\partial\LL$. 
Likewise we need notation for restrictions of the Schr\"odinger operators
to finite cubes with selfadjoint boundary conditions.
We set $H_\per^{L,\rho}= H_0^{L,\rho}+\chi_{\Lambda_L}  V_\per$, $H_\omega^{L,\rho}= H_0^{L,\rho}+\chi_{\Lambda_L} V_\per+\chi_{\Lambda_L} V_\omega$ and similarly for Neumann, Dirichlet and periodic boundary conditions.

Next  we discuss a special choice of mixed boundary conditions introduced by Mezincescu in \cite{Mezincescu-87}. For details see Mezincescu's paper or \cite{KirschW-05}.
Denote by $\psi_1$ the $L^2$-normalized, positive ground state of $H_\per^{1,P}$
and by $\Psi$ its periodic extension on the whole of $\RR^d$.
Then $\psi_L := L^{-d/2} \chi_{\Lambda_L} \Psi$ is the normalized ground state of $H_\per^{L,P}$.
Since $\Psi$ is continuously differentiable and strictly positive (see e.g.~\cite{Simon-82c}), we may define	
$\rho_\Psi(x):=-\frac{n\cdot\nabla\Psi(x)}{\Psi(x)}$.
 We will use the notaton  $H_\per^{L,M}=H_\per^{L,\rho_\Psi}$ and $H_\omega^{L,M}=H_\omega^{L,\rho_\Psi}$ and refer to the corresponding
 boundary conditions as \textsl{Mezincescu boundary conditions}.

 We denote the eigenvalues of the operator $H_\omega^{L,X}$ for $X\in\{D,N,M,P\}$ by  
 \bea
 E_1(H_\omega^{L,X})\leq E_2(H_\omega^{L,X}) \leq\ldots\leq E_n(H_\omega^{L,X})\leq\ldots
 \eea
 with the convention that we repeat eigenvalues according to their multiplicity. 
We also define
 \bea
 N(E,H_\omega^{L,X})~=~\#\{n\mid E_n(H_\omega^{L,X})\leq E\}\;.
 \eea
It is well known that the IDS can be obtained as a macroscopic limit of normalized eigenvalue counting functions
 \bea
 N(E)~=~\lim_{L\to\infty}\,\frac{1}{L^d} N(E,H_\omega^{L,X})
 \eea
 for $X\in\{D,N,M,P\}$. The equality holds for almost all $\omega \in \Omega$ and all energies $E$ where the function $N$ is continuous.
 The following observation of Mezincescu (see \cite{Mezincescu-87} or \cite{KirschW-05}) will be  crucial for our analysis:
\be\label{e-brack}
 N(E)~=~\sup_{L\in \NN}\,\frac{1}{L^d} \,\EE\l(N(E,H_\omega^{L,D})\r)~=~\inf_{L\in \NN}\,\frac{1}{L^d}\, \EE\l(N(E,H_\omega^{L,M})\r)
\ee
Equation \eqref{e-brack} is a so called \emph{bracketing} result. It allows us to estimate the integrated density of states from above
using Mezincescu boundary conditions and from below using Dirichlet boundary conditions.
Another important feature of Mezincescu boundary conditions is the relation
 \bea
 E_1(H_\per^{L,M})~=~E_1(H_\per^{L,P})~=~\inf\sigma(H_\per)~=\inf\Sigma
 \eea
 for all $L$. These relations follow from the following facts: Note 
that $\Psi$ is a positive distributional solution of $H_\per \Psi = E_1(H_\per^{L,P})\Psi$ on all of $\RR^d$.
Thus the Allegretto-Piepenbrink Theorem tells us that $E_1(H_\per^{L,P}) \le \inf\sigma(H_\per)$.
On the other hand, by the Floquet-Bloch decomposition we know that  $E_1(H_\per^{L,P})$ is inside of 
$\sigma(H_\per)$, hence it is the infimum of this set. 
Moreover, the function $\psi_L$ is a normalized $L^2$-eigenfunction of $H_\per^{L,M}$.
Again by positivity it follows that it must be the ground state of this operator.
Finally, note that since the random perturbation is monotone it follows  that
$\inf\sigma(H_\per)~\leq\inf\Sigma$. Standard arguments using sequences of approximate eigenfunctions 
(e.g. as in the proof of equation (1.1) in \cite{KirschSS-98a}) show that $\inf\sigma(H_\per)~\in\Sigma$.
 
Using the fact that $\psi_L$ is the normalized $L^2$-ground-state of $H_\per^{L,M}$
Mezincescu generalizes in \cite{Mezincescu-87} the argument of \cite{KirschS-87} and deduces that
 there exists a constant $\epsilon_0 >0$ such that	
\be\label{e-gap}
 E_2(H^{L,M}_\per)-E_1(H^{L,M}_\per)  \ge \epsilon_0 L^{-2}
\ee
for all $L\in \NN$.

\section{Proof of the upper bound}
\label{s-proof}

In this section we prove the upper bound on the IDS. For simplicity of notation we will assume in the sequel
that $E_0=\inf\sigma(H_\per)=\inf\sigma(H_\omega)=0$. This can always be achieved by
adding a constant to the periodic potential $V_\per$.
Our proof follows the strategy of \cite{KirschS-86}, namely
we will make use of \eqref{e-brack} and of the positivity of $V_\omega$ to estimate for arbitrary 
$L\in \NN$
\be\label{e-esti}
N(E)~\leq~\frac{1}{L^d}\,\EE\l(N(E,H_\omega^{L,M})\r)~\leq~\frac{1}{L^d}\,N(E,H_\per^{L,M})\,\PP\l(E_1(H_\omega^{L,M})<E\r)
\ee
Since we are interested in the behavior of the IDS near the spectral bottom $E_0=0$ it will be sufficient to consider only energies $E\in [0,1]$.
For $E\leq 1$ the quantity $\frac{1}{L^d}\,N(E,H_\per^{L,M})$ is bounded by a constant independet of $L\in \NN, \omega \in \Omega$ and $E\in [0,1]$.
So we are left with the task to bound $\PP\l(E_1(H_\omega^{L,M})<E\r)$ from above. To do so, we estimate $E_1(H_\omega^{L,M})$ from below using Temple's inequality.
We will find that for $E\sim L^{-2}$ we have
\bea
\PP\l(E_1(H_\omega^{L,M})<E\r)~\leq~\exp(C\,L^d) \sim~\exp(C'\,E^{-d/2})
\eea
which together with \eqref{e-esti} gives the desired upper bound on $N$.

Now, we give the details of our proof, which is split into five steps.
We start by introducing new random variables $\xi_k$ which correspont to local energy contributions.

\begin{rem}[Mapped random variables]
On any compact set the function $\Psi$ is strictly positive by the Harnack inequality and bounded by subsolution estimates, 
see e.g.~\cite{CyconFKS-87}. Since $\Psi$ is by definition periodic it is in fact 
uniformly bounded away from zero and from above,
\[
0 <c_3:= \inf_{x\in \RR^d} \Psi(x) \le c_4:= \sup_{x\in \RR^d} \Psi(x) <\infty 
\]
We abbreviate by $d \alpha(x)$ the measure $\Psi(x)^2 dx$ on $\RR^d$.

We introduce for a parameter $c_2\le \frac{\epsilon_0 \epsilon_1}{2\,c_4}$,
i.~e.~$ \frac{c_2\,c_4}{\epsilon_1}\le \frac{\epsilon_0}{2} $, the cut-off random variables
\[
\tilde \lambda_k := \min\{\lambda_k, \lambda_-+c_2 L^{-2} \} \ \in \ [\lambda_-, \lambda_- + c_2 L^{-2}]
\]
and the non-linearly mapped random variables
\[
\xi_k(\omega) = \xi(\tl_k) := \int d\alpha(x)  u  (\tl_k(\omega),x-k)
\]
Each $\xi_k$ corresponds to a summand 
$\langle \Psi, u(\tilde \lambda_k, \cdot-k) \Psi\rangle $
of the energy form depending on the random variable $\lambda_k$.
\end{rem}

We derive two estimates which will be later needed for Temple's inequality.
Set $I_L := \LL \cap \ZZ^d$ and 
denote $\tH_\omega:= H_\per + \tV_\omega$ and $\tV_\omega:=\sum_{k\in \ZZ^d}  u(\tl_k(\omega),x-k)$.

\begin{rem}[Analysis of the first moment]
The following quadratic form will play a crucial role in the sequel.
It may be understood as the first moment of the energy in the state $\psi_L$.
\begin{multline} \label{e:first-moment}
\la \psi_L, \tH_\omega^{L,M} \psi_L \ra
= \la \psi_L, H_\per^{L,M} \psi_L \ra + \la \psi_L, \tV_\omega\psi_L \ra
= \la \psi_L, \tV_\omega \psi_L \ra
\\
= L^{-d}  \sum_{k\in I_L} \int d\alpha(x)   u (\tl_k(\omega), x-k)
= L^{-d}  \sum_{k\in I_L} \xi_k(\omega)
\end{multline}
For $c_2 L^{-2} \le \epsilon_2$, i.~e.~$ \tl_0 \le \lambda_- + \epsilon_2$, we have
\bea
\xi(\tl_0) = \int d\alpha(x) u(\tl_0, x) 
&=& \int_{\lambda_-}^{\tl_0}  d\tau \frac{d}{d\tau} \int d\alpha(x) u(\tau, x)
\\
&\le& \frac{\tl_0- \lambda_-}{\epsilon_1}  \, c_4    \quad  \quad \text{by Assumption (iv)}
\\
&\le&  \frac{c_2}{\epsilon_1} \, \frac{1}{L^2}\, c_4	\quad  \quad \text{ by definition of }  \tl_0
\eea
Hence
\[
\la \psi_L, \tH_\omega^{L,M} \psi_L \ra  = L^{-d} \sum_{k\in I_L} \xi_k(\omega)
\le \frac{c_2\, c_4}{\epsilon_1}  \, \frac{1}{L^2}
\]
and thus for  $\nu:= \frac{\epsilon_0}{2L^2} + \la \psi_L, \tH_\omega^{L,M} \psi_L \ra$
we have
\be
\label{e-}
\nu \le \frac{\epsilon_0}{2}  \, \frac{1}{L^2} + \frac{c_2\, c_4}{\epsilon_1}   \, \frac{1}{L^2}
\le \frac{\epsilon_0}{L^2}  	\le E_2(H_\per^{L,M})
\ee
by the choice of $c_2$, the normalization $E_0=0$, and Ineq.~\eqref{e-gap}.
\medskip
\end{rem}

\begin{rem}[Analysis of the second moment]
We will need also an estimate for the second moment.
By the mean value theorem one sees that for some
$\hat\lambda \in [\lambda_-,\lambda]$
\[
 u ^2 (\lambda,x) = u ^2 (\lambda,x) - u ^2 (\lambda_-,x) 
= 2(\lambda-\lambda_-)  u (\hat\lambda,x) \frac{\partial u (\hat\lambda, x)}{\partial\hat\lambda}
\]
By Assumption (iii) 
we have $0 \le  u (\hat\lambda,x) \le  u (\lambda,x)$ and thus
\begin{align*}
 u ^2 (\lambda,x)
\le 2(\lambda-\lambda_-)  u (\lambda,x) \frac{\partial u }{\partial \lambda} (\hat\lambda, x)
\le 2\kappa_1 (\lambda-\lambda_-)  u (\lambda,x)
\end{align*}
by Assumption (ii). 
Hence
\[
\int d\alpha(x)   u^2(\tl_k, x-k)
\le
2 \kappa_1 c_2 L^{-2}\int d\alpha(x)   u  (\tl_k, x-k) = 2 \kappa_1 c_2 L^{-2}\xi_k
\]
and
\begin{equation}\label{e-variance}
\| \tH_\omega^{L,M} \psi_L \|^2
=
L^{-d}  \sum_{k\in I_L} \int d\alpha(x)   u^2(\tl_k, x-k)
\le
2 \kappa_1 c_2 L^{-2}L^{-d} \sum_{k\in I_L}\xi_k
\end{equation}
\end{rem}
\medskip
\begin{rem}[Lower bound for the first eigenvalue]
The next theorem provides us with a lower bound on the first eigenvalue of a random box Hamiltonian.
It is formulated in terms of an empirical average of the random variables $\xi_k$.
To prove it we use Temple's inequality. The bounds on the first and second moment derived
above are used on one hand to show that Temple's inequality is at all applicable, and
on the other hand to insert them into the inequality to obtain an appropriate lower bound.

\begin{thm}
\label{t-lowerB}
Choose $c_2$ small enough such that $c_2 \leq \epsilon_2 L^2 $ and  $4 \kappa_1 c_2 /\epsilon_0 < 1/4  $. Then
\[
 E_1(\tH_\omega^{L,M})  \ge  \frac{3}{4} L^{-d} \sum_{k\in I_L} \xi_k(\omega)
\]
\end{thm}
\begin{proof}
To ensure that  Temple's inequality can be applied
to the operator $\tH_\omega^{L,M}$ and the vector $\psi_L$,
we need to establish a chain of inequalities, see for instance Theorem XIII.5 in \cite{ReedS-78}.
Since $c_2 L^{-2} \le \epsilon_2$
\begin{align*}
0 = E_1(H_\per^{L,M}) &\le  E_1(\tH_\omega^{L,M})  			& \text{ by monotonicity (iii) 
}
\\ & \le  \la \psi_L, \tH_\omega^{L,M} \psi_L \ra	   		& \text{ by the min-max Theorem}
\\ & <    \nu								& \text{ since $\epsilon_0 >0$}
\\ & \le  E_2(H_\per^{L,M}) 						& \text{ by inequality \eqref{e-}} \label{e-nu-le}
\\ & \le  E_2(\tH_\omega^{L,M}) 					& \text{ by monotonicity (iii) 
}
\end{align*}
We have checked the prerequisites for Temple's inequality and may apply it to the operator $\tH_\omega^{L,M}$
and the vector $\psi_L$:
\bea
E_1(\tH_\omega^{L,M})
&\ge& \la \psi_L, \tH_\omega^{L,M} \psi_L \ra - \frac{\| \tH_\omega^{L,M} \psi_L \|^2}{\nu  -\la \psi_L, \tH_\omega^{L,M} \psi_L \ra}
\\
&\ge& \la \psi_L, \tH_\omega^{L,M} \psi_L \ra -
\frac{2 \kappa_1 c_2 L^{-2}L^{-d} \sum_{k\in I_L}\xi_k(\omega)}{\frac{\epsilon_0}{2}  L^{-2}}
\\
&\ge& \la \psi_L, \tH_\omega^{L,M} \psi_L \ra -
\frac{4 \kappa_1 c_2 }{\epsilon_0} L^{-d} \sum_{k\in I_L}\xi_k(\omega)
\eea
Here we used  equation \eqref{e-variance}.
It follows from equation \eqref{e:first-moment} that 
\[
E_1(\tH_\omega^{L,M})
\ge \l(1 - \frac{4 \kappa_1 c_2 }{\epsilon_0} \r) \la \psi_L, \tH_\omega^{L,M} \psi_L \ra
= \l(1 - \frac{4 \kappa_1 c_2 }{\epsilon_0} \r)\,\frac{1}{L^d} \sum_{k\in I_L} \xi_k(\omega)
\]
and thus we have proven the Theorem.
\end{proof}

The theorem in turn implies an estimate on how small most of the random variables $\xi_k, k\in I_L$ must be, 
if the principal eigenvalue of $\tilde H_\omega^L$ is low.

\begin{cor}
Let $4 \kappa_1 c_2 /\epsilon_0 < 1/4  $ and  $\gamma>1$. Then we have
\[
 E_1(\tH_\omega^{L,M}) \le \cE \ \text{ \rm implies } \ \#\{k \in I_L \mid \xi_k < 2\gamma \cE \} > \frac{\gamma-1}{\gamma}L^d
\]
\end{cor}
\begin{proof}
If the conclusion is false then
\[
\#\{k \in I_L \mid \xi_k \ge 2\gamma \cE \} \ge \frac{L^d}{\gamma}
\]
Hence $ \sum_{k\in I_L} \xi_k(\omega)   \ge 2\gamma \cE   \frac{L^d}{\gamma}  = 2 \cE L^d$.   Theorem \ref{t-lowerB} implies 
\[
 E_1(\tH_\omega^{L,M})  \ge  \frac{3}{4} L^{-d} \cdot\, 2 \cE L^d = \frac{3}{2} \cE > E_1(\tH_\omega^{L,M})
\]
which yields a contradiction.
\end{proof}

\end{rem}
\begin{rem}[Large deviation estimate]
Now we have to show that the event
\[
\#\{k \in I_L \mid \xi_k < 2\gamma \cE \} > \frac{\gamma-1}{\gamma}L^d
\]
has an exponentially small probability in the parameter $L^d$.
To this aim we transform back first to the random variables $\tilde \lambda_k, k\in I_L$
and then to $\lambda_k, k\in I_L$.
\begin{lem}
For $\frac{2\gamma}{\epsilon_1 c_3^2}\leq c_7 \leq \epsilon_2/\cE $,
we have
\[
\xi_k \in [0,2\gamma \cE[  \ \text{ implies } \ \tl_k \in [\lambda_-, \lambda_- + c_7 \cE [
\]
\end{lem}

\begin{proof} Assume $\tl_k \ge \lambda_- + c_7 \cE $.
By (iii) and (vi) 
we have for $c_7 \cE \le \epsilon_2$
\bea
\xi_k(\omega)
&\ge& \int d\alpha (x) u(\lambda_-+ c_7 \cE, x) 
\\
&\geq& c_3^2 \int_{\lambda_-}^{\lambda_- + c_7 \cE} d \tau\frac{d}{d\tau} \int d x\,  u(\tau, x)
\geq c_3^2 c_7 \cE \epsilon_1 \ \text{ }
\eea
Since $c_7 \ge \frac{2\gamma}{\epsilon_1 c_3^2}$, it follows $\xi_k \ge 2\gamma \cE$ which is a contradiction.
\end{proof}
\bigskip

Choose $c_7\cE \leq \frac{c_2}{2L^2} $. Then $\tl_k \in [\lambda_-, \lambda_- + c_7 \cE [$
implies $\lambda_k \in [\lambda_-, \lambda_- + c_7 \cE [$.
Thus we have shown that  for $c_7 \ge \frac{2\gamma}{\epsilon_1 c_3^2}$ 
and $c_7 \cE \leq\min\left(  \epsilon_2,\frac{c_2}{2L^2} \right )$ 
\[
E_1(\tH_\omega^{L,M}) \le  \cE  \quad\text{ implies } \quad
\#\{k \in I_L \mid \lambda_k < \lambda_- + c_7 \cE\} > \frac{\gamma-1}{\gamma}L^d
\]
Since $\mu(\{\lambda_-\}) <1$ there exists a $\lambda_* \in ]\lambda_-,\lambda_+[$ such that 
$p:= \mu([\lambda_*,\lambda_+]) \in ]0,1[$. For $L$ large enough we have $\lambda_- + c_7 \cE \leq \lambda_- + \frac{c_2}{2L^2} \le \lambda_*$,
hence 
\begin{align*}
\PP\left( \#\{k \in I_L \mid \lambda_k < \lambda_- + c_7 \cE\} > \frac{\gamma-1}{\gamma}L^d \right)
&\leq
\PP\left( \#\{k \in I_L \mid \lambda_k < \lambda_*\} > \frac{\gamma-1}{\gamma}L^d \right)
\\
&=
\PP\left( \#\{k \in I_L \mid \lambda_k \geq \lambda_*\} < \frac{1}{\gamma}L^d \right)
  \end{align*}
The latter probability is bounded by $\exp( -\frac{1}{2}  p^2 L^d)$ if we choose $\gamma= 2/p$, 
cf.~Theorem 4.2 in \cite{Simon-85b}.
%
%
To conclude the proof of the upper bound we specify the choice 
\[
L := \left\lfloor \sqrt{\frac{c_2}{2c_7 \cE}}\right \rfloor 
\]

\end{rem}

\section{Proof of the lower bound}
\label{s-lower}
Now we derive a lower bound on the IDS for energies above, but close to $E_0=0$.
For this purpose we deduce from  estimate~\eqref{e-brack} and the \v Ceby\v sev inequality 
\bea
N(E)~\geq~\frac{1}{L^d}\,\EE\l(N(E,H_\omega^{L,D})\r)~\geq~\frac{1}{L^d}\,\PP\big(E_1(H_\omega^{L,D})\leq E\big)
\eea
To bound $E_1(H_\omega^{L,D})$ from above we use the following Lemma which can be found in \cite{KirschS-86} and in \cite{KirschW-05}.

\begin{lem}\label{l-lower} 
There are constants $B_1,B_2\in [0, \infty[$ such that
\bea
E_1(H_\omega^{L,D})~\leq~B_1\,\frac{1}{L^d}\,\int_\LL V_\omega(x)\,dx\,+\,B_2\,L^{-2}
\eea
for all $\omega \in \Omega $ and $L\in \NN$.
\end{lem}
Due to Assumption (iv) we know that for $\lambda_-\leq\lambda\leq\lambda_-+\epsilon_2$ we have
$\int_{\RR^d}u(\lambda,x)\,dx\leq(\lambda-\lambda_-)/\epsilon_1$. So, if $\lambda_k-\lambda_-\leq \delta$ for all $k\in I_L$ with $\delta$ small enough, then
\bea
\frac{1}{L^d}\,\int_\LL V_\omega(x)\,dx~\leq~\frac{1}{L^d}\,\sum_{k \in I_L} \frac{\lambda_k(\omega)-\lambda_-}{\epsilon_1}
\leq~
\frac{\delta}{\epsilon_1}
\eea
and consequently 
$
E_1(H_\omega^{L,D})~\leq~B_1\,\frac{\delta}{\epsilon_1}\,+\,B_2\,L^{-2}
\leq  \frac{3}{4}E
$
if we choose $\delta = \frac{\epsilon_1}{2B_1}E$ and $L := \lceil 2 \sqrt{B_2}\, E^{-1/2} \rceil$.
Combining these estimates we obtain
\begin{align*}
N(E)~&\geq~\frac{1}{L^d}\,\PP\l(\lambda_k-\lambda_-\leq  \frac{\epsilon_1}{2B_1} E; \text{for all $k\in I_L$}\r)\\
&\geq~\frac{1}{L^d}\,\PP\l(\lambda_0-\lambda_-\leq  \frac{\epsilon_1}{2B_1} E\r)^{L^d}
\geq~\frac{1}{L^d}\,\alpha' E^{\kappa\, L^d}
\end{align*}
by Assumption (v), where $\alpha' =\alpha (\frac{\epsilon_1E}{2B_1})^{\kappa\, L^d}$.
Since for $E$ small $L^d \le (4^d B_2^{d/2} ) \, E^{-d/2}$, it follows that
\bea
\lim_{E \searrow 0} \frac{\log |\log N(E)|}{\log E} \geq -\frac{d}{2}
\eea

\centerline{Acknowledgements}
We are grateful to an anonymous referee for careful reading of the manuscript and valuable comments.

\def\cprime{$'$}\def\polhk#1{\setbox0=\hbox{#1}{\ooalign{\hidewidth
  \lower1.5ex\hbox{`}\hidewidth\crcr\unhbox0}}}


\end{document}